\PassOptionsToPackage{dvipsnames}{xcolor}
\PassOptionsToPackage{svgnames}{xcolor}
\PassOptionsToPackage{colorlinks=true,urlcolor=Blue!90,linkcolor=Blue, citecolor=Blue}{hyperref}

\documentclass[nonacm]{acmart}

\usepackage{xcolor}
\usepackage[framemethod=TikZ]{mdframed}
\usepackage[colorlinks=true,urlcolor=Blue!90,linkcolor=Blue, citecolor=Blue]{hyperref}

\usepackage{longtable}
\usepackage{listings}
\usepackage{rotating}
\usepackage{tabularx}
\usepackage{caption}
\usepackage{graphicx}
\usepackage{longtable}
\usepackage{pifont}
\usepackage{enumitem}
\usepackage{tabularx}
\usepackage{lipsum} 
\usepackage{algorithmic}
\usepackage{textcomp}
\usepackage{subcaption}
\usepackage{multirow}
\usepackage{booktabs}
\usepackage[subtle]{savetrees}
\usepackage{titlesec}
\usepackage{wrapfig}
\usepackage{setspace}

\titlespacing*{\section}{0pt}{*0.4}{*0.4}
\titlespacing*{\subsection}{0pt}{*0.4}{*0.4}
\titlespacing*{\subsubsection}{0pt}{*0.5}{*0.5}

\AtBeginDocument{%
  \providecommand\BibTeX{{%
    \normalfont B\kern-0.5em{\scshape i\kern-0.25em b}\kern-0.8em\TeX}}}

\makeatletter
\def\@copyrightspace{\relax}
\makeatother

\makeatletter
\def\@ACM@checkaffil{
    \if@ACM@instpresent\else
    \ClassWarningNoLine{\@classname}{No institution present for an affiliation}%
    \fi
    \if@ACM@citypresent\else
    \ClassWarningNoLine{\@classname}{No city present for an affiliation}%
    \fi
    \if@ACM@countrypresent\else
        \ClassWarningNoLine{\@classname}{No country present for an affiliation}%
    \fi
}
\makeatother

\begin{document}

\title{Assessing the Promise and Pitfalls of ChatGPT for Automated Code Generation}

\author{Muhammad Fawad Akbar Khan$^*$}
\email{khan@usu.edu}
\affiliation{%
  \institution{Department of Computer Science, Utah State University}
  \city{Logan}
  \state{Utah}
  \country{USA}
  \postcode{84322}
}

\author{Max	Ramsdell$^*$}
\email{a02237674@usu.edu}
\affiliation{%
  \institution{Department of Computer Science, Utah State University}
  \city{Logan}
  \state{Utah}
  \postcode{84322}
}

\author{Erik Falor}
\email{erik.falor@usu.edu}
\affiliation{%
  \institution{Department of Computer Science, Utah State University}
  \city{Logan}
  \state{Utah}
  \postcode{84322}
}

\author{Hamid Karimi}
\email{hamid.karimi@usu.edu}
\affiliation{%
  \institution{Department of Computer Science, Utah State University}
  \city{Logan}
  \state{Utah}
  \postcode{84322}
}

\begin{abstract}

This paper presents a comprehensive evaluation of the code generation capabilities of ChatGPT, a prominent large language model, compared to human programmers. A novel dataset of 131 code-generation prompts across 5 categories was curated to enable robust analysis. Code solutions were generated by both ChatGPT and humans for all prompts, resulting in 262 code samples. A meticulous manual assessment methodology prioritized evaluating correctness, comprehensibility, and security using 14 established code quality metrics. The key findings reveal ChatGPT's strengths in crafting concise, efficient code with advanced constructs, showcasing strengths in data analysis tasks (93.1\% accuracy) but limitations in visual-graphical challenges. Comparative analysis with human code highlights ChatGPT's inclination towards modular design and superior error handling. Additionally, machine learning models effectively distinguished ChatGPT from human code with up to 88\% accuracy, suggesting detectable coding style disparities. By providing profound insights into ChatGPT's code generation capabilities and limitations through quantitative metrics and qualitative analysis, this study makes valuable contributions toward advancing AI-based programming assistants. The curated dataset and methodology offer a robust foundation for future research in this nascent domain. All data and codes are available on \href{https://github.com/DSAatUSU/ChatGPT-promises-and-pitfalls/}{https://github.com/DSAatUSU/ChatGPT-promises-and-pitfalls/}.
\end{abstract}

\ccsdesc[500]{Computing methodologies~Artificial intelligence}
\ccsdesc[500]{Computing methodologies~Natural language processing}
\ccsdesc[300]{Human-centered computing~Human computer interaction (HCI)}
\ccsdesc[300]{Applied computing~Education}

\keywords{ChatGPT, Human, Code Generation, Software, Code Metrics, Machine Learning, Code Classification}

\maketitle

\begingroup
\renewcommand\thefootnote{*}
\footnotetext{These authors contributed equally to this work.}
\endgroup

\section{Introduction}

Artificial intelligence (AI) has demonstrated significant promise in revolutionizing the automation of various intricate tasks within the domain of software engineering. Notably, cutting-edge methodologies in code generation now harness AI models to autonomously create complete or partial software programs based on either natural language descriptions or specific code inputs~\cite{tian2023chatgpt}. The recent rise of large-scale language models (LLMs) has garnered considerable attention, not only in broader society but also within the realm of software engineering, pushing AI-driven automation to unprecedented levels~\cite{chen2021evaluating, devlin2019bert, tian2020evaluating}. These LLMs, pre-trained on extensive datasets encompassing both source code and natural language, have exhibited remarkable proficiency in comprehending code structures and generating code or textual content. The progress propelled by LLMs has significantly bolstered the efficacy of automated methods for tackling a range of challenges in software engineering, including code generation~\cite{chakraborty2022natgen,dakhel2023github, fried2023incoder}, program rectification~\cite{dakhel2023github,jain2021jigsaw,jiang2023impact}, and code summarization~\cite{ahmed2022fewshot, hu2020deep}.

Generative Pre-trained Transformer (GPT) models~\cite{brown2020language}, specifically ChatGPT~\cite{openai-chatgpt-blog} created by OpenAI, is an advanced language model tailored for conversational applications such as question-answering and code generation. Built upon the GPT-3.5 series architecture, ChatGPT boasts an impressive 175 billion parameters and has been fine-tuned using reinforcement learning with human feedback. This extensive training enables ChatGPT to produce responses that closely resemble human language based on context comprehension. The ongoing conversation history has garnered substantial interest within the software engineering community. This interest is primarily attributed to ChatGPT's reported capability in realizing a longstanding aspiration in software engineering: the automatic repair of software with minimal human intervention~\cite{sobania2023analysis}. These reported outcomes suggest that ChatGPT holds transformative potential for the field, indicating a promising future for LLM-driven software engineering and AI programming assistant tools. However, further research is imperative to precisely delineate the extent of LLM capabilities specifically for generating programs.

This paper focuses on conducting an extensive and systematic evaluation of code generated by ChatGPT, with a particular emphasis on assessing its correctness, comprehensibility, and security. We have chosen ChatGPT, a prominent and widely recognized LLM, to serve as a representative example of LLMs. Many studies that focus on generating programs with ChatGPT tend to assess its performance using outdated benchmark data available prior to 2022--See Section~\ref{sec:related work}. This outdated data might have inadvertently influenced the training data for ChatGPT. Additionally, none of these studies conduct a comparative analysis between ChatGPT-generated code and human-written code, effectively highlighting the limitations of using the ChatGPT model for code generation. This potential bias in experimental design raises concerns about the applicability of the reported results to new and unforeseen challenges. Furthermore, the existing literature lacks a comprehensive examination of ChatGPT's capabilities in code generation, emphasizing the need for further investigation in this area. By conducting this comprehensive analysis, we aim to contribute to advancing ChatGPT-based code generation techniques, potentially enhancing the broader field of AI and LLM-based code generation.  In this study, we aim to provide insightful answers to the following fundamental questions:

\begin{enumerate}
    \item [\ding{73}] \textbf{Functional Accuracy}: Can ChatGPT outperform humans in generating functional and highly accurate code?
    \item [\ding{73}]  \textbf{Code Understandability}: Is ChatGPT capable of producing code that is more understandable than that of humans?
    \item [\ding{73}]  \textbf{Code Security}: Can ChatGPT demonstrate superior capabilities in generating code that is more secure compared to human-generated code?
    \item [\ding{73}]  \textbf{ChatGPT code detection}: Is it possible to develop a prediction model that can reliably distinguish between ChatGPT-generated and human-generated code reliably, achieving a significant level of accuracy?
\end{enumerate}

Our contributions that followed in this study are summarized below.
\vspace{-0.1cm}
\begin{itemize}[leftmargin=0.8cm]
    \item [\ding{112}] \textbf{Constructing a New Dataset:} We curate a diverse dataset of 131 prompts across five categories and 25 subcategories featuring human-written code. This dataset forms the foundation for a robust comparative analysis, assessing the efficacy of code-generation algorithms against human-coded solutions.
    
    \item [\ding{112}]  \textbf{Comprehensive Code Evaluation:} Using ChatGPT3.5 Turbo, a widely recognized LLM, we systematically evaluate generated code, prioritizing correctness, understandability, maintainability, and security. This evaluation employs 14 interpretable code quality metrics.
    
    \item [\ding{112}]  \textbf{Comparative Analysis with Human Code:} We conduct a comparative analysis between ChatGPT-generated and human-written code across 131 prompts in five categories, revealing both limitations and strengths in ChatGPT's code generation across different categories of code.
    
      \item [\ding{112}]  \textbf{Critical Analysis of Prompts:} Our research emphasizes the influence of prompt quality on ChatGPT's code generation capabilities through case studies, providing key considerations for prompt design, an emerging skill in LLMs.
    
     \item [\ding{112}]  \textbf{ChatGPT Code Detection:} Using introduced features, we develop machine learning (ML) models to classify ChatGPT code versus human code. Additionally, we conduct a reliability test to ensure the model's generalizability. To the best of our knowledge, this is the first ML ChatGPT code detection model. 
\end{itemize}

\subsection{Broader Impacts and Significance}

This research provides significant broader impacts, offering valuable insights into the emerging capabilities of AI-based code generation tools like ChatGPT. As large language models continue rapid advancements, it is imperative that the software engineering and computer science education communities develop a nuanced understanding of their promise and limitations. By systematically evaluating ChatGPT's code generation proficiency across diverse domains, our work helps delineate the current state of its abilities compared to human programmers. 

Our findings have particular relevance to the field of learning analytics. With the rise of intelligent tutoring systems and AI teaching assistants, it is critical to precisely determine their competencies and shortcomings in automating programming education. Our analysis offers data-driven guidance on ChatGPT's reliability in grading programming assignments, providing feedback, and assisting students. Educators can utilize our insights to make informed decisions on incorporating ChatGPT while considering its flaws, like struggles with visual and advanced tasks.

Furthermore, our introduced dataset and comparative analysis methodology provide a rigorous foundation for future learning analytics research assessing AI tutors. As schools increasingly blend online tools with in-person instruction, robust evaluations of AI's pedagogical effectiveness are imperative. Our work pioneers techniques to gauge automated programming tutors' capacities across languages, problem categories, and metrics like correctness. Overall, by spurring prudent AI adoption while advancing evaluation methods, this research delivers significant broader impacts for the learning analytics community.

\section{Related Work}
\label{sec:related work}

The synergy of Educational Data Mining (EDM) and Learning Analytics (LA) has manifested a profound capacity to decipher and employ educational data for enhancing learning and teaching methods. These fields intricately interlace to parse educational environments and tailor pedagogical strategies to individual learning needs~\cite{karimi2019roadmap,karimi2021automatic,karimi2020online,solanki2023leveraging,farokhi2023enhancing}. The assimilation of sophisticated machine learning techniques in EDM and LA has not only refined predictive models for student success but has also provided a granular understanding of educational interactions within online platforms~\cite{karimi2019roadmap}.

In parallel, the emergence and evolution of GPT-based Large Language Models (LLMs) have garnered substantial attention for their prowess in Natural Language Processing (NLP) and Code Generation~\cite{sobania2023analysis}. These LLMs, such as ChatGPT, have demonstrated exceptional performance in generating human-like text, compelling the research community to delve into their application within educational settings~\cite{liu2023your,liu2023no}. Fan et al.'s examination of automated program repair (APR) techniques like Codex underscores the potential for these technologies to remedy code generation limitations, especially within the context of coding problem-solving platforms like LeetCode~\cite{fan2023automated}. Xia et al.~\cite{pearce2023examining} conducted a comprehensive study involving the direct application of nine state-of-the-art Language Models (LLMs) for Automated Program Repair (APR). Their evaluation encompassed various approaches for utilizing LLMs in APR, such as entire patch fixes, code chunk fixes, and single-line fixes. Hendricks et al.~\cite{hendrycks2021measuring} introduced a benchmark for Python programming problems called "craftAPPS". They evaluated the code generation performance of several GPT-based variant models by fine-tuning them with the craftAPPS dataset. Dong et al.~\cite{dong2023self} introduced the idea of a software development lifecycle and put forth a self-collaboration framework. This framework utilizes distinct ChatGPT conversations in various roles, such as analyst, developer, and tester, to collaborate in the code generation process. Liu et al.~\cite{liu2023refining} analyzed various code quality issues associated with ChatGPT-based code generation. However, most of these studies utilize a publically available dataset, for example, LeetCode problems \cite{LeetCode} and CWE (Common Weakness Enumeration) scenarios (CWE’s code scenarios) as provided in~\cite{pearce2022asleep}. The challenge with existing code datasets lies in their lack of customization to effectively evaluate ChatGPT's code generation capabilities. For instance, these datasets may neglect scenarios involving the creation of visually intensive, graphical, or drawing-oriented programs. Furthermore, the effectiveness of the GPT for code generation is poorly understood, and the generation performance could be heavily influenced by the choice of prompt~\cite{liu2023improving}. Therefore, using an outdated prompt dataset such as (e.g., OpenAICookbook~\cite{openaicookbook}, and PromptBase~\cite{mastropaolo2021studying}) may not be an effective option. Prompts should be custom engineering, offering sufficient information to ChatGPT while leveraging its dialog ability. Additionally, conducting preliminary analyses of the prompts can yield deeper insights into ChatGPT's code-generation abilities, potentially leading to more suitable prompts for an enhanced dataset. The datasets for analysis should be thoroughly tailored to encompass a comprehensive spectrum of programming aspects, spanning various categories and subcategories. This approach can effectively uncover vulnerabilities and limitations within ChatGPT's code generation capabilities.

Against this backdrop, our research stands out. We introduce a meticulously crafted prompt dataset spanning five distinct categories, offering a richer evaluation canvas than generic benchmarks. By juxtaposing ChatGPT code with human-generated code, we attain a deeper comprehension. Our use of 14 code quality metrics and novel machine learning models for ChatGPT code detection sets a new standard in the domain. Our holistic approach, emphasizing rigorous benchmarks and thorough methodologies, brings forth novel insights into ChatGPT's code generation capabilities. Consequently, our contributions lay a robust groundwork that can spur further innovations in AI-driven programming tools.

\section{Data Collection and Curation Process}
To maximize ChatGPT's performance, we conducted our study using Python3, a highly expressive programming language. A study done by Zhijie Liu et al. verified that ChatGPT is better at generating Python3 code in terms of understandability, functionality, and security metrics compared to 4 other languages C++, C, JavaScript, and Java~\cite{liu2023no}. 

\begin{figure}[ht]
    \centering
    \includegraphics[width=1\linewidth]{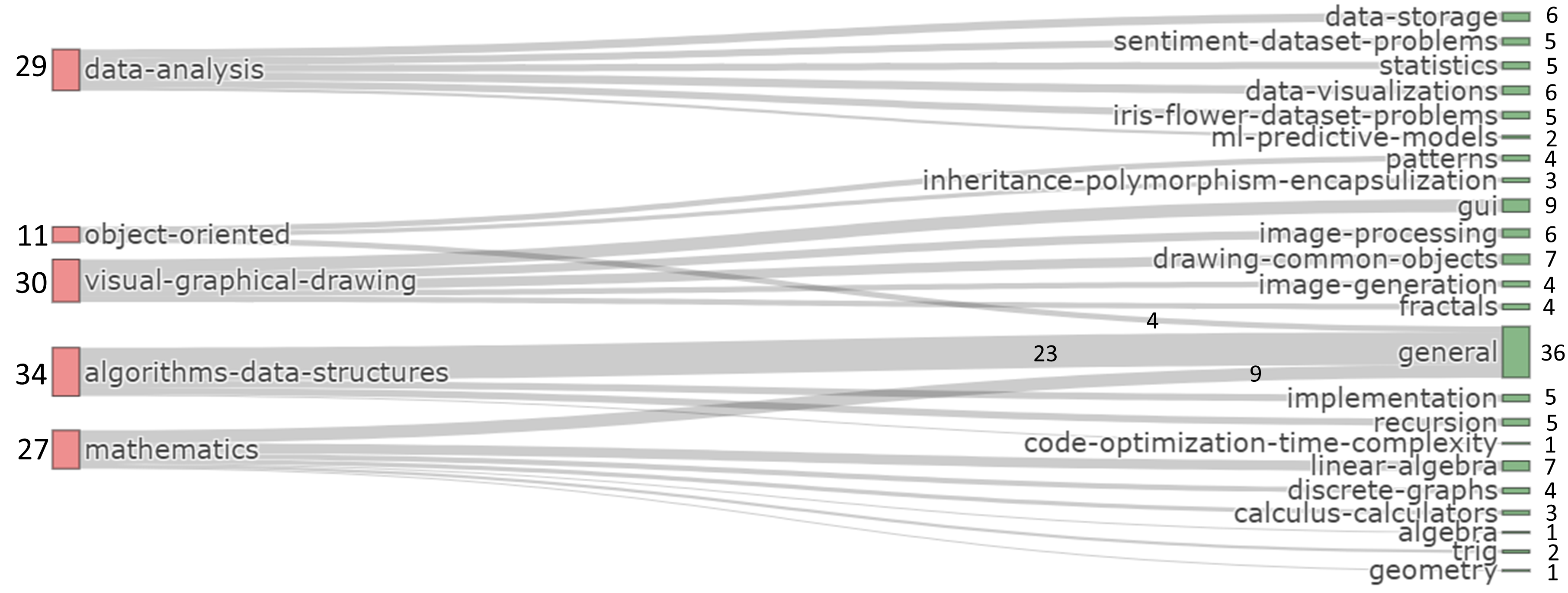}
      \vspace{-0.5cm}
    \caption{ Categories and subcategories for which data and prompts were collected. The numbers show the total number of prompts collected for that category or subcategory.}
     \label{fig:sankey-diagram}

\end{figure}
\subsection{Programming Prompt Conceptualization}
To pinpoint the focal area for prompt collection, we conducted a preliminary investigation, thoroughly exploring different facets of ChatGPT's code generation capabilities. This investigation utilized assignments from the ``Introduction to Python Programming" course (CS1) at a major US public university as ChatGPT's initial prompts. Tailored for computer science undergraduates without prior programming experience, these assignments covered diverse programming concepts, including visual and drawing problems, algorithms, data structures, loops, and object-oriented programming. The insights from the preliminary prompts guided our decision-making on categories and subcategories essential for a comprehensive analysis of ChatGPT's code-generation capabilities, as depicted in Figure~\ref{fig:sankey-diagram}. To construct this framework, we gathered prompts from diverse online platforms, including Github, Medium, GeekforGeeks, and others.

\begin{itemize}
    \item [\ding{234}] \textbf{Algorithms and Data Structures (ADS):} This category features prompts of varying difficulty levels in \textit{algorithms and data structures}, evaluating ChatGPT's ability to devise solutions for sorting, searching, recursion, optimization, arrays, linked lists, trees, and graphs. This category assesses its understanding of mathematics and computer science concepts.
    
    \item [\ding{234}] \textbf{Data Analysis (DA):} This category includes prompts of diverse complexity in \textit{data analysis}, assessing ChatGPT's capacity to generate code for data cleaning, manipulation, visualization, and statistical analysis. The goal is to evaluate ChatGPT's proficiency in addressing real-world data tasks.

    \item [\ding{234}] \textbf{Mathematics (M):} This category incorporates prompts of diverse complexity in \textit{mathematics}, covering basic geometry, trigonometry, arithmetic, algebra, calculus, and advanced topics. This category evaluates ChatGPT's ability to generate code solutions for a range of mathematical problems.

    \item [\ding{234}] \textbf{Object Oriented (OO):} This category includes prompts of varying complexity within \textit{object-oriented programming}, addressing tasks related to class design, inheritance, polymorphism, encapsulation, and design patterns. The aim is to assess ChatGPT's ability to generate code adhering to object-oriented principles.

    \item [\ding{234}] \textbf{Visual Graphical Drawing (VGD):} The last category focuses on ChatGPT's ability to generate code for \textit{visual patterns, drawing, and graphical} challenges. This category includes tasks ranging from turtle graphics patterns to visualizing directions, complex designs, GUIs, pixel art, and image processing. Given ChatGPT's potential challenges in this domain (described in Section~\ref{sec:experiments}), evaluating its code generation proficiency here is crucial.
    
\end{itemize}

\begin{figure}[ht]
    \centering
    \includegraphics[width=\columnwidth]{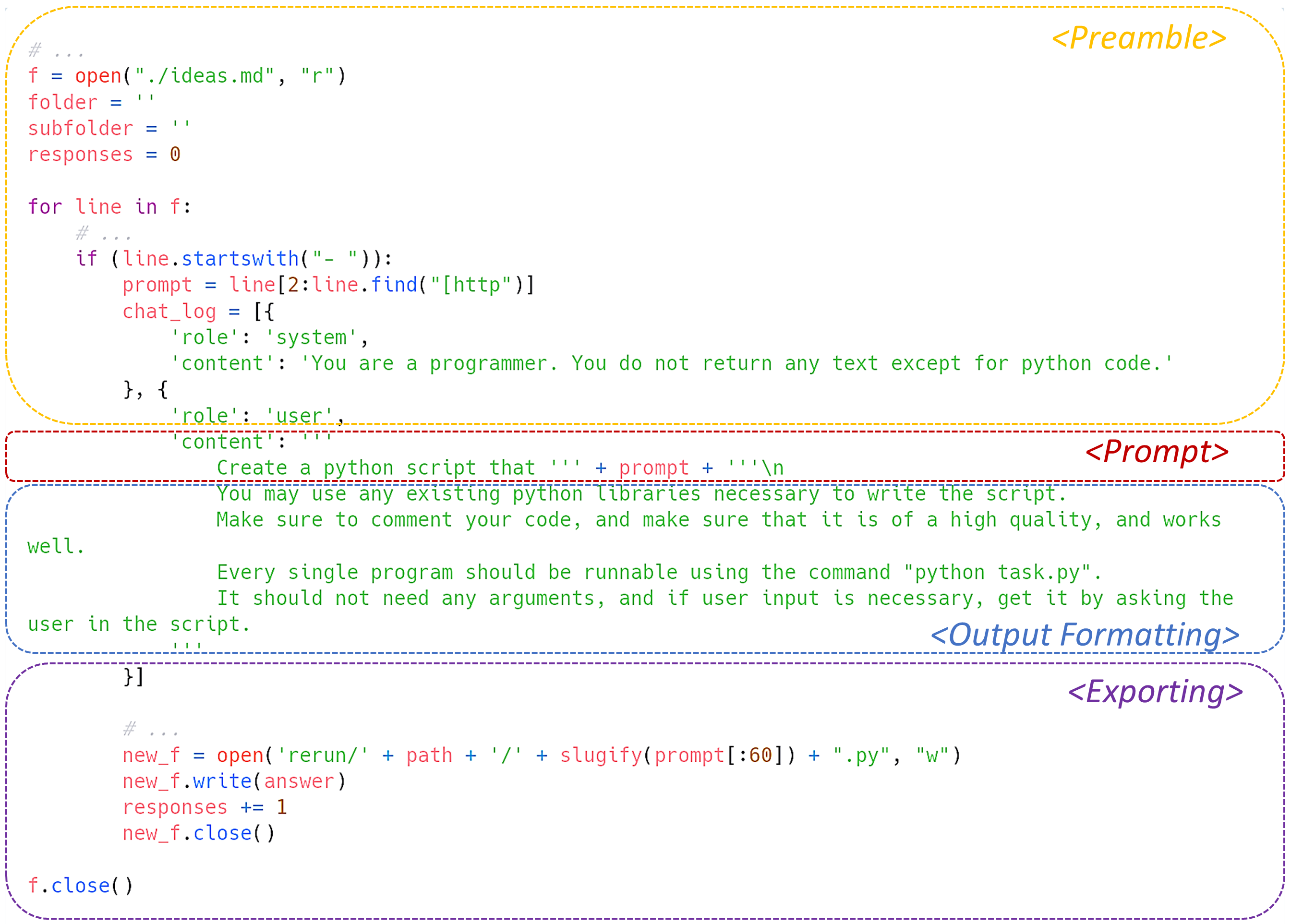}
      \vspace{-0.5cm}
    \caption{Code generation script for automatically processing each prompt and exporting code into specified directories.}
    \label{fig:prompt-figure}
\end{figure}
\subsection{Collection of Prompts}
In total, 131 prompts were collected. To ensure consistency, a prompt template was designed with four essential components \textit{<Preamble>, <Prompt>, <Output Formatting> and <Exporting>}. Figure~\ref{fig:prompt-figure} shows the script used to collect the codes for each prompt. Out of the 131 total prompts, approximately 60 prompts were collected from online sources, and 76 prompts were engineered specifically for this study, mainly in the \textit{visual-graphica-drawing} and \textit{algorithms-data-structures} categories accounting for {23\%} and {26\%} of the total prompts, respectively. 

\subsection{ChatGPT Code Generation}
To streamline the process, we developed a shell script that harnessed ChatGPT Turbo 3.5 API keys (July 2023) to generate code solutions for each of the 131 prompts automatically. To ensure the quality and functionality of the generated code from the prompts, we initiated the process by manually assessing a few sample codes, iterative refining the prompts, and verifying ChatGPT's performance. Following the successful validation of our automated process, we efficiently generated code solutions for all 131 prompts, resulting in a total of 131 code files. Subsequently, we organized the obtained code into separate \textit{.py} files and meticulously structured them within nested directories to ease subsequent analysis.

 \subsection{Human Code Collection}
To create the human-coded dataset for all 131 prompts, we engaged six computer science major individuals, each responsible for selecting prompts based on their preference. They were instructed to produce code solutions independently in their unique coding styles, with strict guidance not to utilize AI assistance tools, including ChatGPT, GitHub Copilot, or Kite. However, they could access online resources to collect ideas to generate the code. Approximately 80 prompts were meticulously crafted through manual coding in this manner. In order to diversify the dataset to encompass a broader range of coding styles, an additional approximately 51 codes were sourced from various online platforms. These codes were meticulously curated to ensure their authors were human programmers. Particular emphasis was placed on retrieving code snippets from the most recent (after \textit{Sept 2021}) or non-public sources whenever feasible to guarantee that the code had not been incorporated into the training set of ChatGPT 3.5.

\section{Methodology}
Utilizing a dataset of 262 code samples (131 from ChatGPT and 131 from humans), we performed a collaborative effort for a manual evaluation of each code. The focus was primarily on assessing functionality.  A customized test suite was meticulously designed for each prompt. This approach facilitated a comprehensive examination of the code's capabilities and helped identify potential limitations in ChatGPT's coding capabilities. For codes resulting in compilation errors, runtime errors, or incorrect output, a corrective process ensued, spanning up to 10 rounds. With each successive round, task-related information was incrementally provided within the prompts to guide the code generation process. Codes that remained unsuccessful even after the 10-round correction process were categorized as incorrect. For human-generated code, we did not impose any time or round limits, allowing for the completion and execution of all codes, consistently yielding correct outputs. For ChatGPT-generated code, the functional correctness of codes was manually assessed independently by the authors. In the subsequent sections, we elaborate on the code metrics utilized in our analysis and elucidate the methodology employed for training the machine learning algorithm.

\subsection{Code Analytical Metrics}
\label{subsection:metrics}
To comprehensively assess ChatGPT's code compared to human-generated code, we employed 14 well-established programming metrics commonly utilized in the research literature. These metrics serve a dual purpose, enabling us to not only gain insights into the coding style of ChatGPT and humans but also facilitate in-depth examinations of specific coding behaviors, thus enhancing our observations. This approach allows for a thorough evaluation of functionality, understandability, and security.
\begin{enumerate}
    \item \textbf{Cyclomatic Complexity:} Cyclomatic Complexity quantifies the number of linearly independent paths in code, offering a pure complexity measurement. Lower complexity, indicative of fewer branches, aligns with the goal of achieving comprehensive code coverage. This metric is frequently employed to determine the number of paths necessary for full testing coverage. Additionally, it has been observed that Cyclomatic Complexity tends to correlate with Source Lines of Code (SLoC) strongly and may share similar predictive capabilities.\\
    \textbf{Halstead Metrics:} Halstead metrics are derived from the code's count of operators (e.g., '+' or 'int') and operands (e.g., variable names or numbers). These metrics aim to quantify the "physical" properties of code, similar to how physical matter is characterized by mass and volume.
        \item \textbf{Halstead Difficulty}: Measures the code's readability and understanding.
        \item \textbf{Halstead Effort}: Estimates the effort required to write the code.
          \item \textbf{Halstead Volume}: Reflects the program's size, including operator and operand counts.
          \item \textbf{Halstead Time}: Predicts the time needed to develop the program.
          \item \textbf{Halstead Bugs}: Estimates potential bug count, aiding in debugging efforts.
      \item \textbf{Source Lines of Code (SLoC):} The physical number of lines of code. Essentially, it takes the total lines of code for a file and removes whitespace lines and comment lines.
      \item \textbf{Logical Lines of Code (LLoC):} The number of statements in a program. For example, a line that has a print statement after an if statement on the same line would have 1 SLoC and 2 LLoC.
      \item \textbf{Difference of SLoC LLoC:} Since the difference between SLoC and LLoC can provide insight into the quality of code. Higher quality code will have lower difference values. Therefore we add that as a metric.
      \item \textbf{Number of Lines:} Total number of lines in the code. Counting code, new lines, and comments. This is different from LLoC and SLoc.
      \item \textbf{Number of Comments:} The count of total number of comments in the code. This counts the comments by counting the start of each comment.
      \item \textbf{Number of Functions:} Count of the number of functions used in code.
      \item \textbf{Number of Classes:} Count of the number of classes used in code.
      \item \textbf{Maintainability Index:} The Maintainability Index is a software metric designed to measure how maintainable and comprehensible a software system is. It considers factors such as code size, complexity, and coupling, providing a numerical score that reflects the ease with which developers can understand, modify, and maintain the codebase. A higher Maintainability Index indicates better code maintainability.
      
\end{enumerate}
Using these code metrics, we embark on an in-depth analysis to identify the nuance difference between human-generated and ChatGPT-generated code using scientific visualization.

\subsection{ChatGPT Code Detection}
Utilizing 14 code metrics, we trained seven machine learning algorithms—including Decision Trees (DT), Random Forest (RF), and K-Nearest Neighbors (KNN)—to discern between ChatGPT and human code. The typical training procedure involved:
\begin{itemize}
    \item [\ding{113}] \textbf{Train-Test Split:} An 80-20\% ratio ensured 210 codes in the training set and 52 in the test set from both ChatGPT and humans.
    \item [\ding{113}]\textbf{Hyperparameters Tuning:} Parameters were determined through grid search based on literature and experimentation.
    \item [\ding{113}]\textbf{Training:} Each algorithm was trained with optimized hyperparameters, and performance was gauged using standard metrics. To mitigate randomness, models were trained multiple times with varied seeds.
\end{itemize}

\subsubsection{Reliability Tests.}
We conducted reliability tests to evaluate the classifiers' real-world applicability:
\begin{itemize}
    \item [\ding{113}]\textbf{Train-Test Split Ratio Test:} We experimented with train-test ratios from 10\% to 100\% in 5\% increments.
    \item [\ding{113}]\textbf{Per-category Performance Test:} Checked model performance across categories to prevent overfitting to any single category.
    \item [\ding{113}]\textbf{Gaussian Noise Test:} The classifiers' resilience was tested by introducing Gaussian noise to the dataset.
\end{itemize}

\subsubsection{Feature Analysis.}
The 14 features across 262 data instances were analyzed to discern coding style variations. Two methods were adopted:
\begin{itemize}
    \item [\ding{113}]\textbf{Random Forest-based Approach:} Evaluated importance through impurity decrease metrics.
    \item [\ding{113}]\textbf{Feature Permutation-based Technique:} Assessed feature relevance by modifying performance metrics after feature permutation.
\end{itemize}
By averaging results from both methods over 1000 iterations, we gained a robust understanding of feature importance, balancing the strengths and weaknesses of each method.

\begin{figure}[ht]
  \centering
  \includegraphics[width=1.\linewidth]{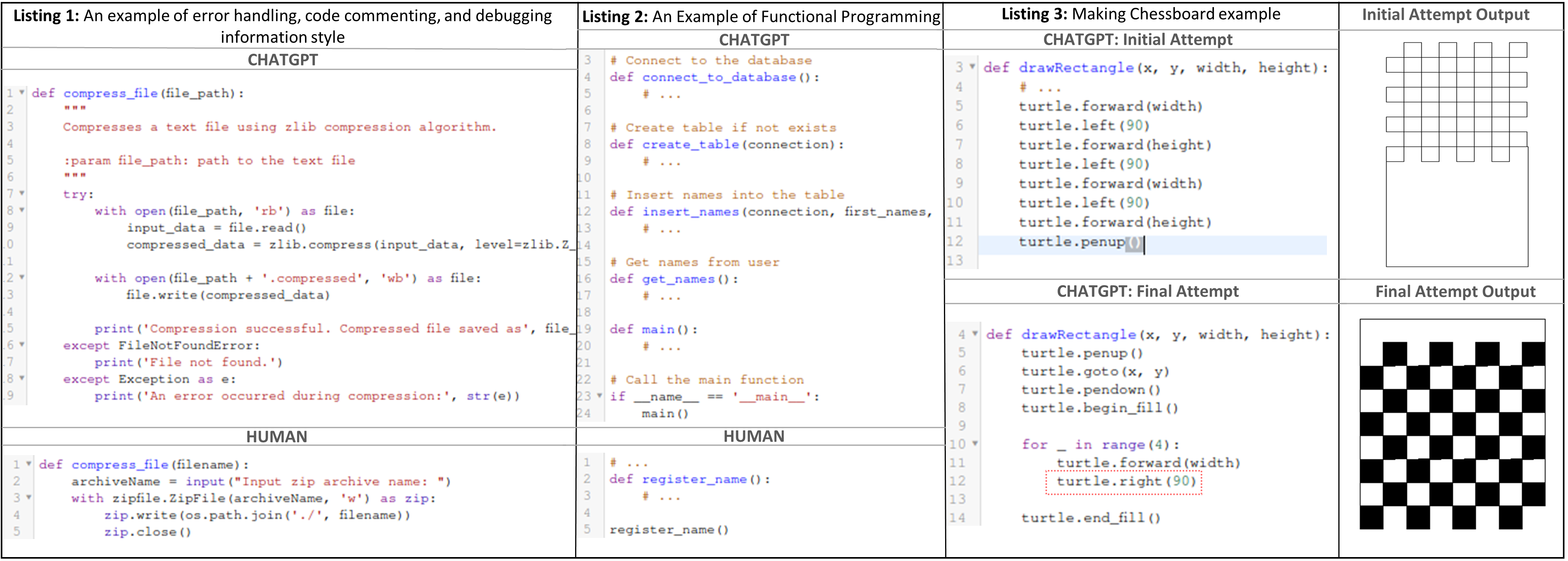}
  \vspace{-0.5cm}
  \caption{Codes and outputs for the presented case studies}
  \label{fig:listings-figure}
\end{figure}
\vspace{-0.4cm}
\section{Experimental Results}
\label{sec:experiments}
In this section, we present experimental results delineating ChatGPT deep code analysis and prediction.  

\subsection{Functional Accuracy}
After the assessment of 262 codes for functional accuracy, the results for the 131 prompts were categorized into three classes:

\begin{enumerate}
    \item [\ding{172}] \textbf{Correct and Compilable:} These codes were not only accurate in fulfilling the task description but also provided the required output across the test suite.
    \item [\ding{173}]  \textbf{Incorrect but Compilable:} This category includes compilable codes that produced incorrect output. It also includes instances of codes with infinite loops resulting in a "time limit exceeded" condition.
    \item [\ding{174}]  \textbf{Incorrect and Uncompilable:} Codes falling into this category were deemed uncompilable due to various reasons, such as compile errors, type errors, time limit exceeded errors or runtime errors. 
\end{enumerate}

\subsubsection*{\textbf{Correct and Compilable}}
Among the accurately generated prompts, a substantial majority were classified under the Data Analysis category, with 93.1\% of the prompts being correct. This highlights ChatGPT's robust comprehension of concepts related to predictive modeling, data storage, statistics, and data visualizations. Out of the total 16 failures, 7 (23.3\%), 3 (8.8\%), 3 (11.11\%), 2 (6.8\%), and 1 (9\%) occurred in the VGD, ADS, M, DA, and OO categories, respectively.

\subsubsection*{\textbf{Incorrect and Compilable.}}

In another scenario, we observed that overloading ChatGPT with excessive details and imposing numerous conditions could adversely impact its code generation capabilities. This leads to ChatGPT desperately trying to solve the problem by generating code that is compilable but with incorrect output. We provide insights about this category through the following two case studies. 

  \vspace{0.2cm}

  \begin{mdframed}[linewidth=1pt,linecolor=black,backgroundcolor=Blue!3,roundcorner=5pt]
  \begin{center}
    \textbf{Case Study 1: Drawing Chessboard}
\end{center}

    In our preliminary investigation, ChatGPT was tasked with replicating a Python script assignment where CS students at a large public university in the US created a chessboard using the Turtle module. The assignment was in a 3-page PDF file with detailed instructions about the task implementation and execution. Despite providing detailed prompts with instructions on how to implement and execute the task, ChatGPT consistently produced incorrect but compilable code, struggling with nuances like the chessboard border placement missing tiles and misalignment of tiles and the box. Interestingly, simplifying the prompt to a concise instruction, "\textit{Create Python code to generate an 8x8 chessboard using the Turtle Python package}," resulted in accurate code on the first attempt. This suggests that, akin to humans, ChatGPT excels at smaller and straightforward tasks. Notably, the performance improvement parallels the success of human students given a similar concise prompt for a smaller chessboard task, underscoring the impact of task framing on language model performance.

 \textbf{ChatGPT Directional Dyslexia:} When addressing issues in the ChatGPT output, it became apparent that while ChatGPT correctly followed the prescribed steps, it encountered challenges in aligning the tiles within the chessboard bounding box or rectangle. A chessboard consists of tiles inside a rectangular box. To create a tile, we utilized the commands \texttt{pen.right()} and \texttt{pen.forward()} repeated four times. However, a problem arose as ChatGPT struggled to discern how using \texttt{pen.right()} or \texttt{pen.left()} to construct the tiles (black squares) would impact their alignment within the box--See Listing 3 Code and Output in Figure~\ref{fig:listings-figure}. Notably, the use of \texttt{pen.right()} resulted in the first row of tiles being positioned below and outside the chessboard, and this alignment issue was rectified by employing \texttt{pen.left()}.
  \end{mdframed}

\vspace{0.5cm}

  \begin{mdframed}[linewidth=1pt,linecolor=black,backgroundcolor=Blue!3,roundcorner=5pt]

\begin{center}
    \textbf{Case Study 2: Generating Sprite}
\end{center}

Contextual meaning is crucial for ChatGPT to provide accurate output. For instance, a prompt instructing the machine to create a script generating "Sprites" for video games resulted in a script producing a flat image. Recognizing the misunderstanding, we modified the term "Sprite" to "pixel art image," leading to a script generating images resembling static on an old television—technically correct as a form of "random pixel art image." This highlights the significance of providing contextual framing for prompts, as ChatGPT may struggle with words carrying double meanings without such context, unlike humans familiar with the associated concepts.
\end{mdframed}

\subsubsection*{\textbf{Incorrect and Uncompilable.}} 
Non-compilable codes, marked by compile or runtime errors persisting even after 10 rounds of incremental correction attempts, revealed ChatGPT's struggle in adapting to outdated methods and functions. The model's lack of awareness of changes beyond September 2021 renders it a liability for coding tasks with newer contexts or requirements involving recently released packages. Additionally, the generated code may employ outdated methods, potentially leading to slower performance and suboptimal memory management. To provide more detail about the nature of this category, we present the following case study. 
\vspace{0.3cm}
  \begin{mdframed}[linewidth=1pt,linecolor=black,backgroundcolor=Blue!3,roundcorner=5pt]
  \begin{center}
      \textbf{Case Study 3: \texttt{load\_boston()}}
  \end{center}

ChatGPT relies on training data up to September 2021, potentially leading to outdated knowledge of packages. In an example, when tasked to create a script using "sklearn" for a predictive model, ChatGPT chose the "Boston Housing Dataset," unaware it had been deprecated since sklearn's 1.2 update in December 2022. This limitation means ChatGPT might generate code with deprecated or nearly deprecated features, necessitating caution. In contrast, human-written code, while not always the latest, can offer more recent, relevant, and reliable solutions. Despite initial failures, corrective adjustments, such as changing \texttt{load\_boston()} to \texttt{fetch\_california\_housing()}, were deemed correct, showcasing a nuanced evaluation approach.
\end{mdframed}

\noindent {\textbf{Conclusion.}} In conclusion, our evaluation encompassed three key aspects of ChatGPT's coding capabilities. In generating \textit{Correct and Compilable} code, ChatGPT exhibited proficiency with a commendable 93.1\% accuracy, notably emphasizing modularity through a higher number of functions. However, when tackling \textit{"Incorrect but Compilable"} codes, the model faced challenges in tasks requiring advanced visual imagination and multi-layered problem-solving, showcasing limitations in creativity and optimal code generation. Notably, ChatGPT lacks the ability to leverage visual feedback for quick error detection, hindering accuracy in drawing and visual problem-solving. Lastly, the examination of \textit{"Incorrect and Uncompilable"} codes highlighted the model's struggle with outdated methods, emphasizing the importance of continual model updates to adapt to evolving coding practices. While ChatGPT's limitations were evident, successful corrective adjustments demonstrated the model's potential for improvement with targeted interventions, offering valuable insights for future advancements in natural language programming models.

\subsection{Code Understandability, Maintainability and Security}
To assess ChatGPT's code qualities compared to humans, we use metrics box plots and radar plots (Figures.~\ref{fig:box_whiskers_plot} and \ref{fig:radar_plot}) of introduced metrics in Section~\ref{subsection:metrics}. These visualizations offer insights into the understandability, maintainability, and security of both ChatGPT and human codes. Analyzing programming styles, our observations are as follows:

\begin{figure}[ht]
  \centering
  \begin{minipage}{.52\textwidth}
    \centering
    \includegraphics[width=\linewidth]{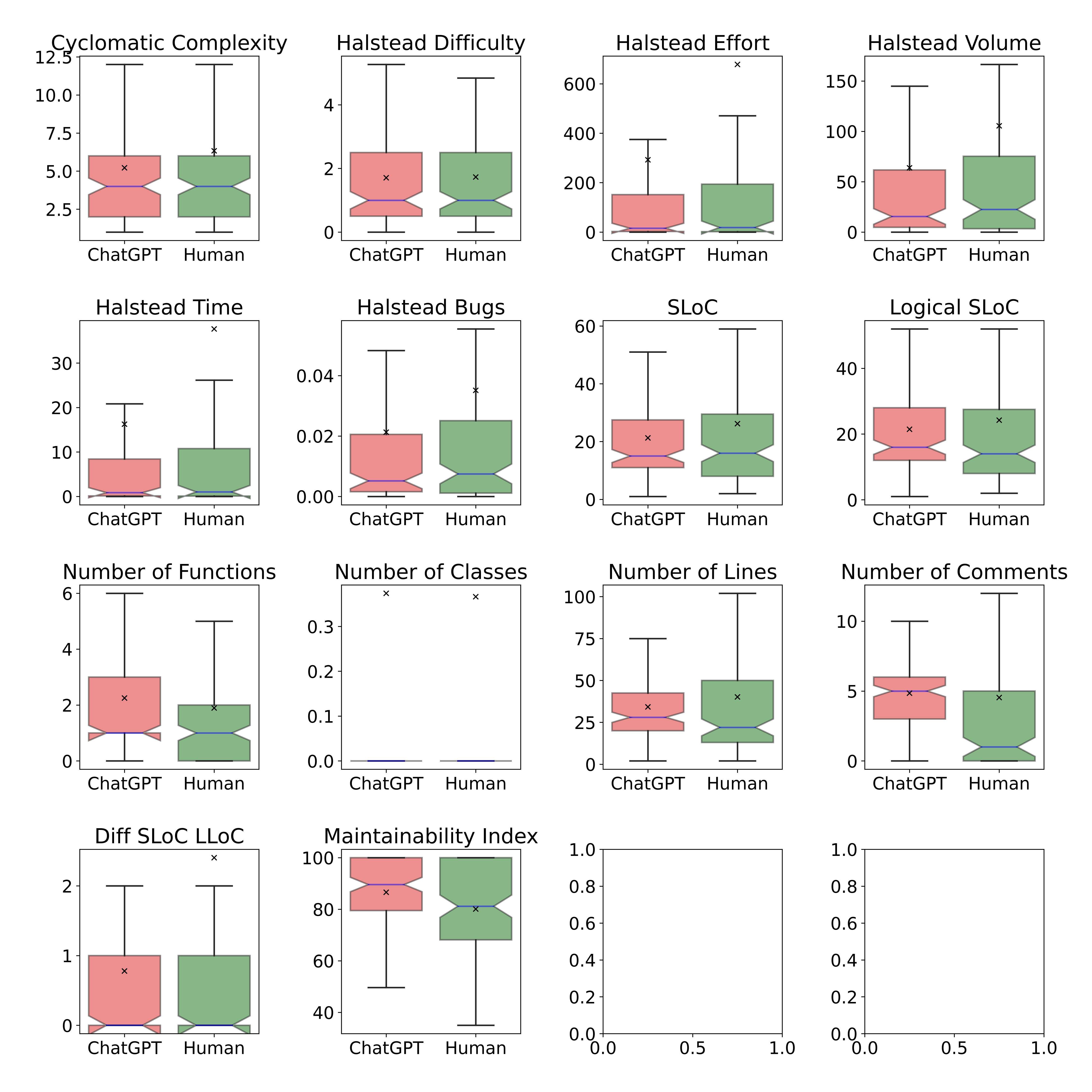}
    \caption{This box and whisker plot illustrates key values of analytical metrics: Median (line within the box), Mean ('x' symbol), Minimum (lower whisker), Maximum (upper whisker), Lower Quartile (bottom of the box), and Upper Quartile (top of the box) for each metric}
    \label{fig:box_whiskers_plot}
  \end{minipage}\hfill
  \begin{minipage}{.48\textwidth}
    \centering
    \includegraphics[width=\linewidth]{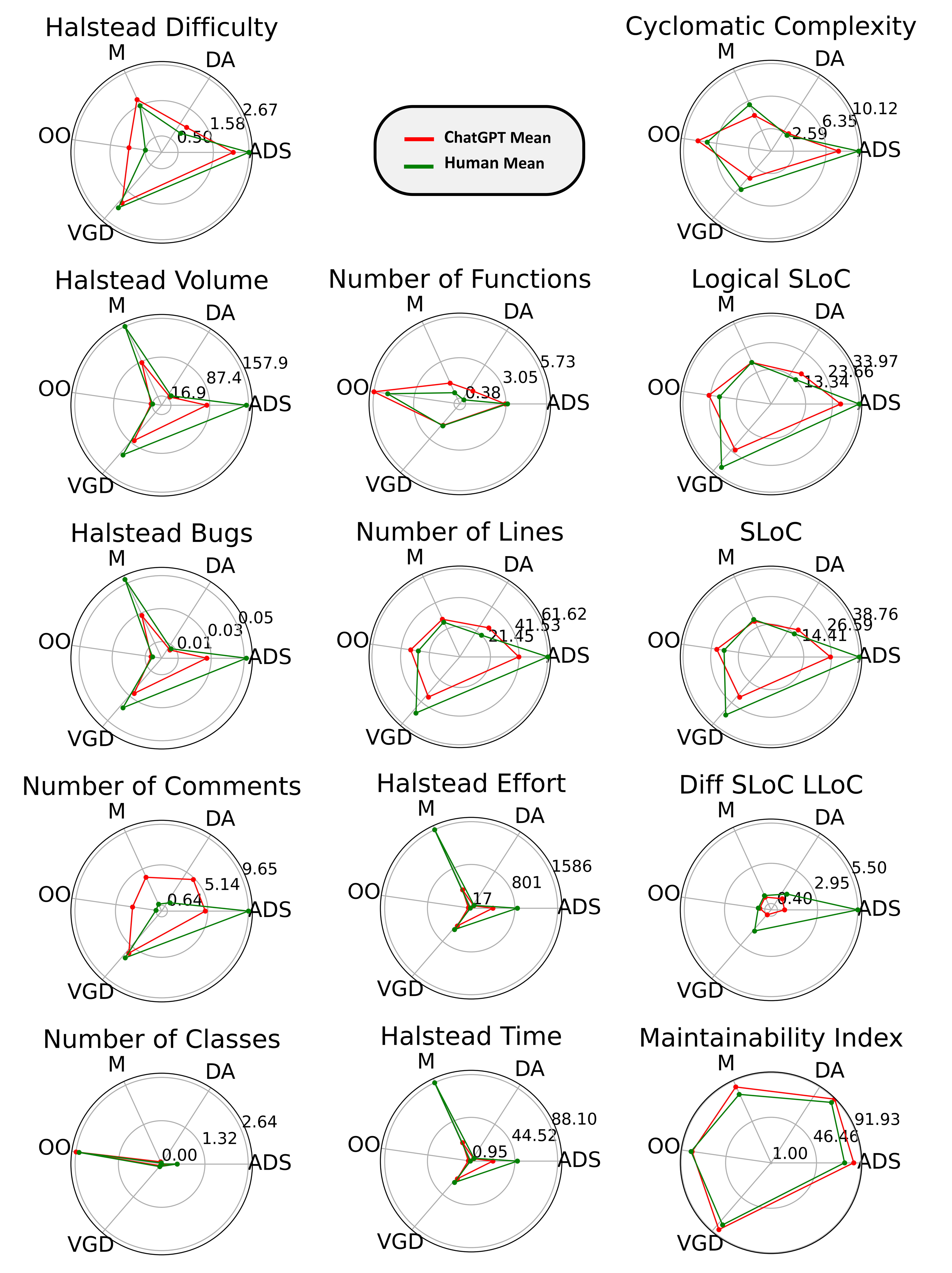}
    \caption{This radar plot displays the mean for each metric across categories, providing a comprehensive view of ChatGPT and human coding styles across categories}
    \label{fig:radar_plot}
  \end{minipage}
\end{figure}

\ding{112} \textbf{Functional Programming:} The plots highlight ChatGPT's significant use of functions, particularly in the M and ADS categories, showcasing its inclination towards functional programming. In contrast, humans tend to minimize unnecessary functional usage. This distinction is evident in the spider plot, with a notable contrast in the number of functions used by ChatGPT and humans, especially in the M and ADS categories. However, this difference is less apparent in the OO, DA, and VGD categories, suggesting a nuanced response influenced by the emphasis on functional programming in the prompts. Additionally, as programs increase in size and complexity, humans typically introduce more functions. The strategic use of functional programming enhances ChatGPT's coding efficiency, leading to fewer Logical Lines of Code (LLoC) and a higher maintainability index.

\ding{112} \textbf{Lines of Code and Comments} Significant disparities in the distribution of the Number of Lines of Code and Comments are evident across the entire dataset and within each category. Generally, humans tend to produce more lines of code and fewer comments, exhibiting sporadic behavior. In contrast, ChatGPT consistently generates concise and efficient code with succinct comments, typically not exceeding one line. This distinction is more pronounced in the VGD and ADS categories, containing the most challenging problems. Notably, ChatGPT tends to provide more comments for OO, M, and DA categories while offering fewer comments for VGD and ADS categories. This pattern arises from humans' tendency to avoid extensive comments and longer codes, particularly when confronted with the complexity of challenges in VGD and ADS categories.

\ding{112} \textbf{Complexity of Code (LoC):} ChatGPT consistently generates code with higher cyclomatic complexity, particularly in the challenging M, ADS, and VGD categories, showcasing its ability to produce intricate solutions. This is evident in the relatively lower disparity between physical and logical lines of code, highlighting ChatGPT's efficiency in crafting concise codes using advanced Python concepts such as list comprehensions, functional programming, inheritance, polymorphism, generators, iterators, decorators, and sophisticated data structures. Notably, its strength lies in compact code generation across most categories, except for VGD and ADS, where performance was subpar, leading to more incorrect solutions. In comparison, human programmers may avoid these advanced techniques due to perceived complexity. A cross-category analysis reveals ChatGPT's weaknesses in visual, graphical, or drawing problems.

\ding{112} \textbf{Halstead Metrics:} Additionally, we evaluated the quality of the generated code using Halstead metrics, including difficulty, effort, volume, and time. Humans showed an anomalously high standard deviation of difficulty, volume, effort, and bugs for the mathematics category, which means some prompts were poorly executed by humans compared to ChatGPT. Overall, the ChatGPT code had lower means for all Halstead metrics than humans, specifically for M, ADS, and VGD categories. This showed that ChatGPT was able to provide efficient code for these categories. However, it also performed poorly regarding functional accuracy in the VGD and ADS categories.

\ding{112}  \textbf{Maintainability Index:} Furthermore, ChatGPT's code exhibits enhanced maintainability across all categories from a maintainability and security perspective. This is also characterized by fewer bugs measured by the Halstead Bugs metric and the implementation of high-quality error-handling techniques. These techniques encompass robust \textit{exception handling}, the adoption of Testing and Test-Driven Development (TDD) practices, a strong emphasis on functional programming principles, and proficient memory management strategies. 

We accompany these observations with two case studies:

\vspace{0.3cm}
  \begin{mdframed}[linewidth=1pt,linecolor=black,backgroundcolor=Blue!3,roundcorner=5pt]
  \begin{center}
      \textbf{Case Study 4: Error Handling, Code Commenting, and Debugging Information}
  \end{center}
 In many of the prompts, ChatGPT was seen to perform excellent error handling, commenting, and debugging output, as shown in Listing 1 in Figure~\ref{fig:listings-figure}. In that example, fairly indicative of the differences between many of the scripts, it is easy to see that the error handling code is more comprehensive, there are more comments, and the debugging output is of higher quality in the ChatGPT written script than the human one.
Focusing on error handling, ChatGPT wrote excellent error handling code, taking into account not just a general case but also the specific \textit{FileNotFoundError} exception as well. The human code was written without any error handling.
Also note that if you remove the error handling, comments, and debugging output, the GPT code would be a single line longer than the human code. 
In this situation, ChatGPT prioritizes error handling and debugging output to make the code more useable and understandable despite not having been specifically instructed to do so. We can also see it outperforming the human code in the number of comments explaining the code and can note that the debugging output could be considered its own form of comment, further promoting the idea that ChatGPT is proficient at writing code that follows general best practices for writing high quality and understandable code.
\end{mdframed}

\vspace{0.3cm}
  \begin{mdframed}[linewidth=1pt,linecolor=black,backgroundcolor=Blue!3,roundcorner=5pt]
  \begin{center}

\textbf{Case Study 5: Functional Programming}
\end{center}
It was seen that ChatGPT used functions at a significantly greater rate than humans. The code in Listing 2 of Figure~\ref{fig:listings-figure} showcases this use of functions. It is easy to see that the program generated by ChatGPT has 5 different functions, each performing a different task, while the human code has a single function.
We can see here that GPT used functions to promote understandability and reusability in its code, which significantly contrasts with the human code in this case. This was often the case in most of the ChatGPT codes, while humans tend to code function only when necessary.
\end{mdframed}

\textbf{Conclusion.} Comparing code understandability, maintainability, and security between ChatGPT and humans, metric distribution plots reveal that ChatGPT tends to generate code with higher cyclomatic complexity, excelling in crafting concise code using advanced Python concepts. ChatGPT displays proficiency in list comprehension, functional programming, and sophisticated data structures. Evaluation with Halstead metrics consistently indicates that ChatGPT produces higher-quality code, exemplified by lower difficulty and time scores, while maintaining lower effort and volume scores compared to human-generated code. From a maintainability and security perspective, ChatGPT exhibits enhanced qualities, showcasing robust error handling, commenting, and debugging output. Despite potential biases in the instructions given, the observations emphasize the ChatGPT's capability to produce code with advanced constructs and quality practices.

\subsection{ChatGPT Code Detection using Machine Learning}
This section presents our findings concerning hyperparameter tuning for the seven classification algorithms, their classification performance, and their resilience during reliability testing.

\begin{table}[htpb]
\centering
\vspace{-0.2cm}
\caption{Performance of various classification models in predicting ChatGPT code}
\vspace{-0.3cm}
\label{table:classification-results}
\resizebox{0.6\textwidth}{!}{%
\begin{tabular}{@{}ccccccc@{}}
\toprule
\textbf{Model} & \textbf{Class} & \textbf{Precision} & \textbf{Recall} & \textbf{F1} & \textbf{Accuracy} & \textbf{Weighted F1} \\ \midrule

\multirow{2}{*}{RF} & ChatGPT & 83\% & 92\% & 87\% & \multirow{2}{*}{87\%} & \multirow{2}{*}{0.8649} \\
 & Human & 91\% & 81\% & 86\% & & \\ \hline

\multirow{2}{*}{DT} & ChatGPT & 83\% & 96\% & 89\% & \multirow{2}{*}{88\%} & \multirow{2}{*}{0.8839} \\
 & Human & 95\% & 81\% & 88\% & & \\ \hline

\multirow{2}{*}{RUSBoost} & ChatGPT & 79\% & 85\% & 81\% & \multirow{2}{*}{81\%} & \multirow{2}{*}{0.8050} \\
 & Human & 83\% & 77\% & 80\% & & \\ \hline

\multirow{2}{*}{GaussianNB} & ChatGPT & 51\% & 96\% & 67\% & \multirow{2}{*}{52\%} & \multirow{2}{*}{0.40} \\
 & Human & 67\% & 8\% & 14\% & & \\ \hline

\multirow{2}{*}{MLP} & ChatGPT & 51\% & 85\% & 64\% & \multirow{2}{*}{52\%} & \multirow{2}{*}{0.4617} \\
 & Human & 56\% & 19\% & 29\% & & \\ \hline

\multirow{2}{*}{XGB} & ChatGPT & 69\% & 85\% & 76\% & \multirow{2}{*}{73\%} & \multirow{2}{*}{0.7308} \\
 & Human & 80\% & 62\% & 70\% & & \\ \hline

\multirow{2}{*}{KNN} & ChatGPT & 61\% & 85\% & 71\% & \multirow{2}{*}{65\%} & \multirow{2}{*}{0.6415} \\
 & Human & 75\% & 46\% & 57\% & & \\ \bottomrule
\end{tabular}
}
\end{table}
\vspace{-0.3cm}
\subsubsection{Code Detection Performance.} Table~\ref{table:classification-results} demonstrates the performance of ChatGPT code detection using different machine learning models and across different measures.  The Decision Trees (DT) algorithm exhibited exceptional performance with an accuracy of 88\%, closely followed by the Random Forest (RF) algorithm, which achieved an accuracy of 87\%. Since the dataset is perfectly balanced, the F1 and Weighted F1 scores align closely. It is intriguing to observe that RUSBoost, typically employed for imbalanced datasets, delivered outstanding results with an accuracy of 81\%. RF, DT, and RUSBoost outperformed all other algorithms by a substantial margin.

\subsubsection{Reliability Tests.}
The results of the three reliability tests provided insight into the model's generalizability. For this analysis, we only used the top three best-performing algorithms, i.e., RF, DT, and RUSBoost algorithms. Tests were performed using the best hyperparameter.  The following observations were made from each test.

\begin{figure}[ht]
  \centering
  \begin{subfigure}{0.30\linewidth}
    \includegraphics[width=\linewidth]{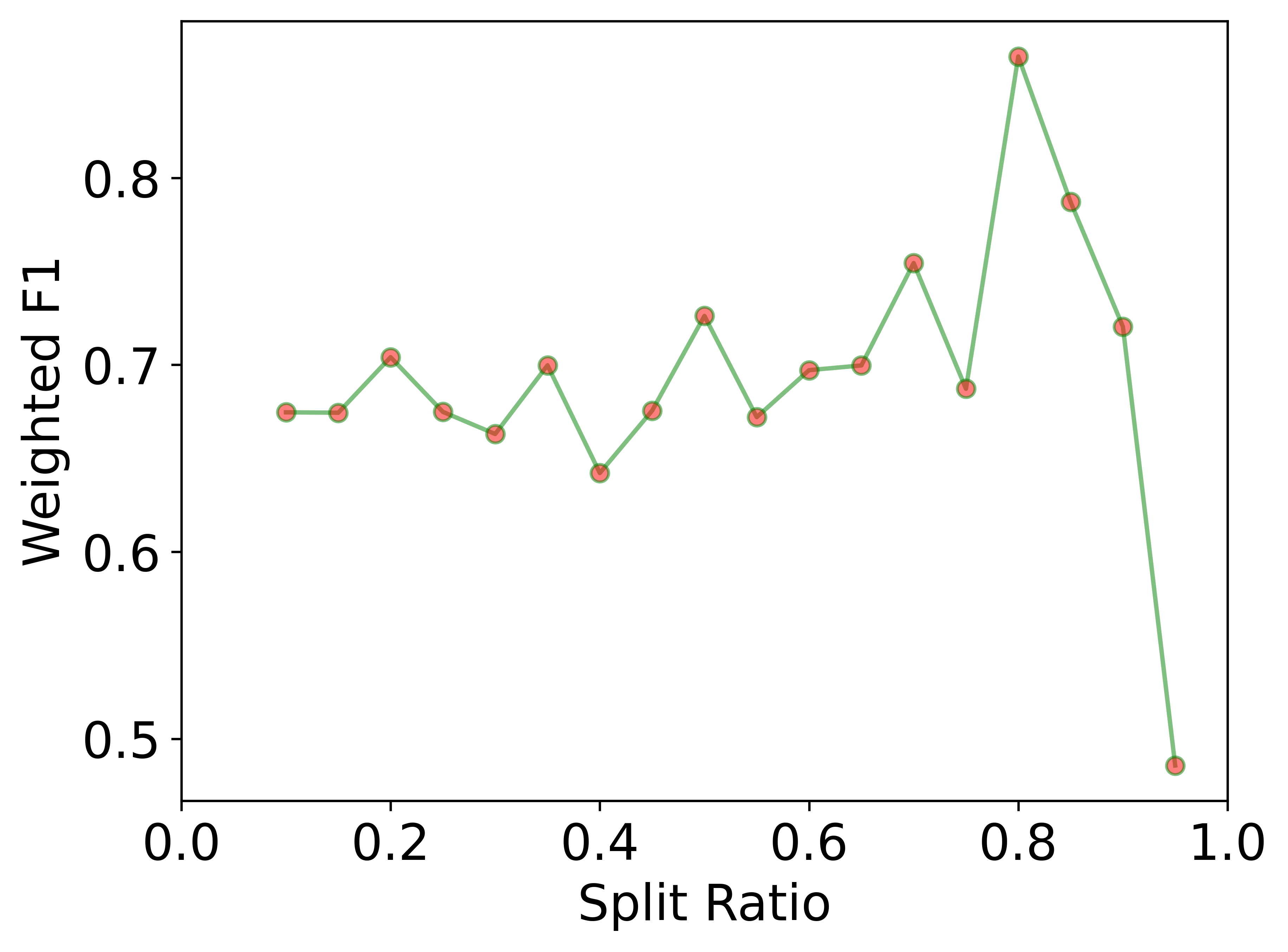}
    \caption{Random Forest}
    \label{fig:train_split_rf}
  \end{subfigure}
  \hfill
  \begin{subfigure}{0.30\linewidth}
    \includegraphics[width=\linewidth]{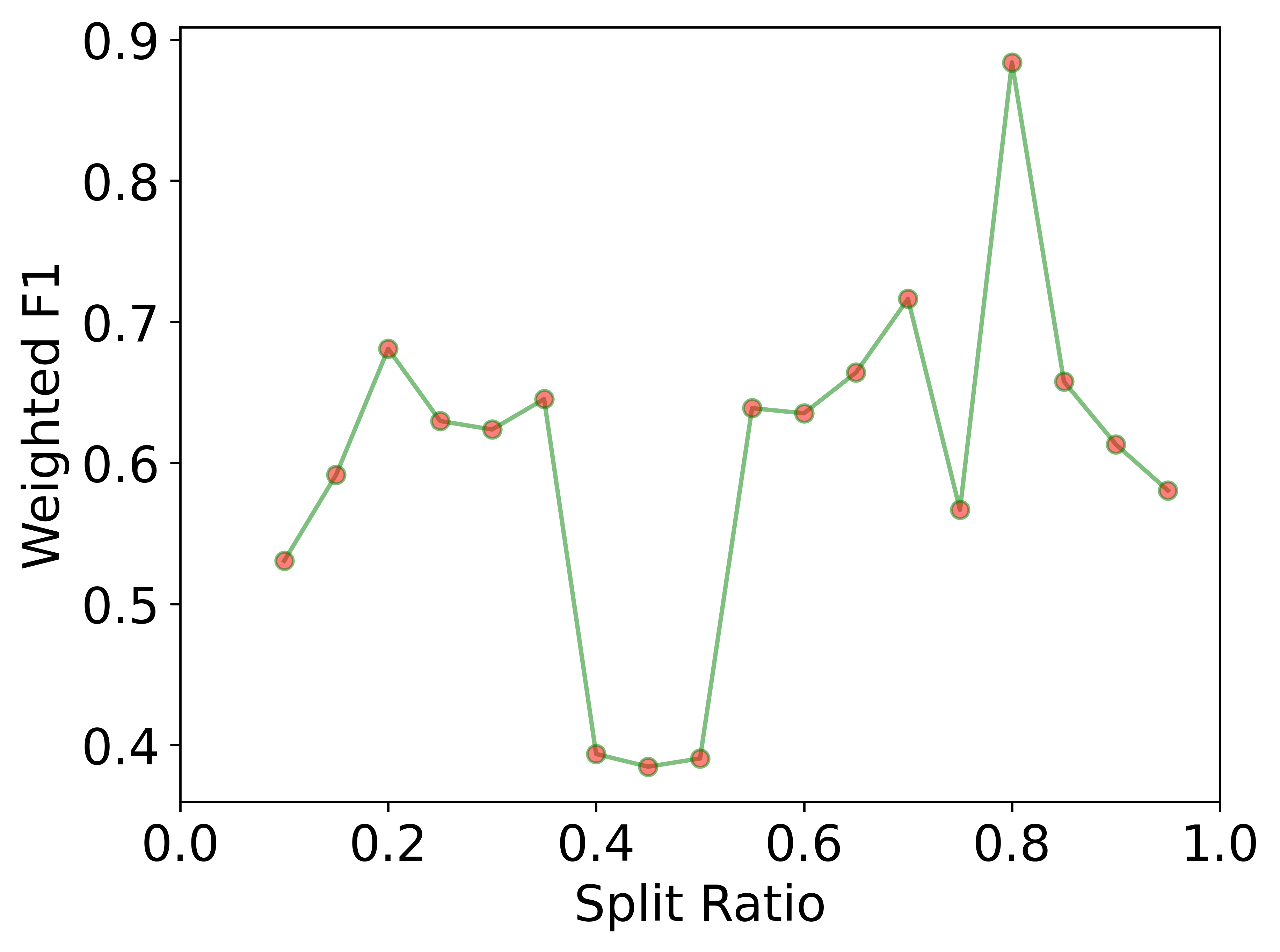}
    \caption{Decision Tree}
    \label{fig:train_split_dt}
  \end{subfigure}
  \hfill
  \begin{subfigure}{0.30\linewidth}
    \includegraphics[width=\linewidth]{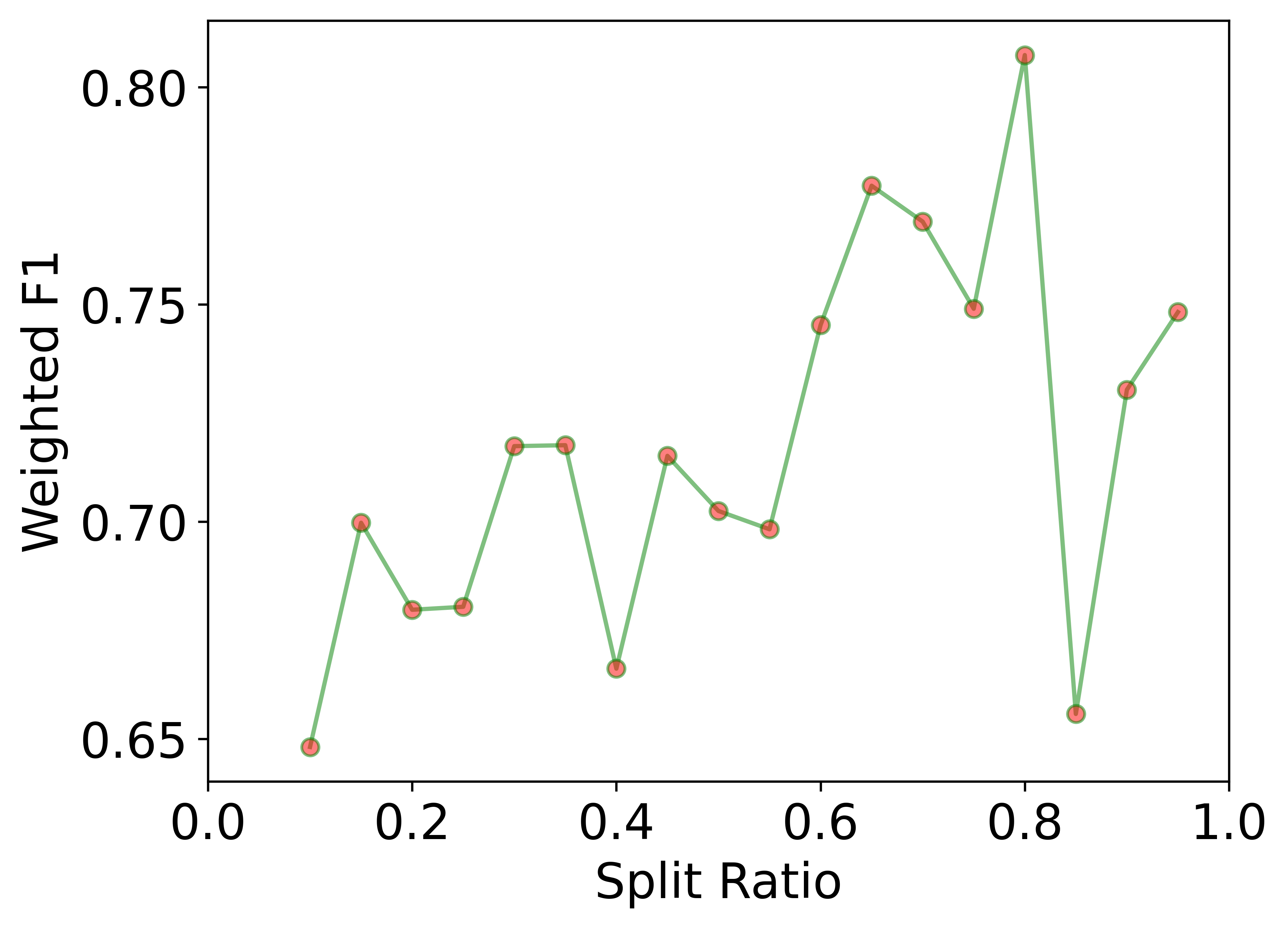}
    \caption{RUSBoost}
    \label{fig:train_split_rusboost}
  \end{subfigure}
  \vspace{-0.3cm}
  \caption{ML model performance on the test set for different train split size (x-axis)}
  \label{fig:train_split_plots}
\end{figure}
\ding{112}  \textbf{Train-Test Split Ratio Test.} Examining the impact of varied train set ratios on model performance, we noted distinct behaviors among the three models. Random Forest (RF) demonstrated remarkable stability, achieving the highest 87\% F1 Score at an 80\% train split size as seen in Figure~\ref{fig:train_split_plots}. Conversely, the Decision Tree (DT) model showcased the highest accuracy of 88\% but exhibited inconsistent performance with increasing train split size, potentially attributed to random dataset sampling after selecting the train split ratio. Notably, a significant performance decline at the 40-50\% train split size range raised concerns about DT's generalizability. Random Under-Sampling with Boosting (RUSBoost) displayed similar variation, suggesting a risk of overfitting. Interestingly, all models experienced a sharp drop in performance after an 80\% train split size, possibly due to a limited number of examples in the test set.
   
 \begin{figure}[ht]
  \centering
  \begin{subfigure}{0.32\linewidth}
    \includegraphics[width=\linewidth]{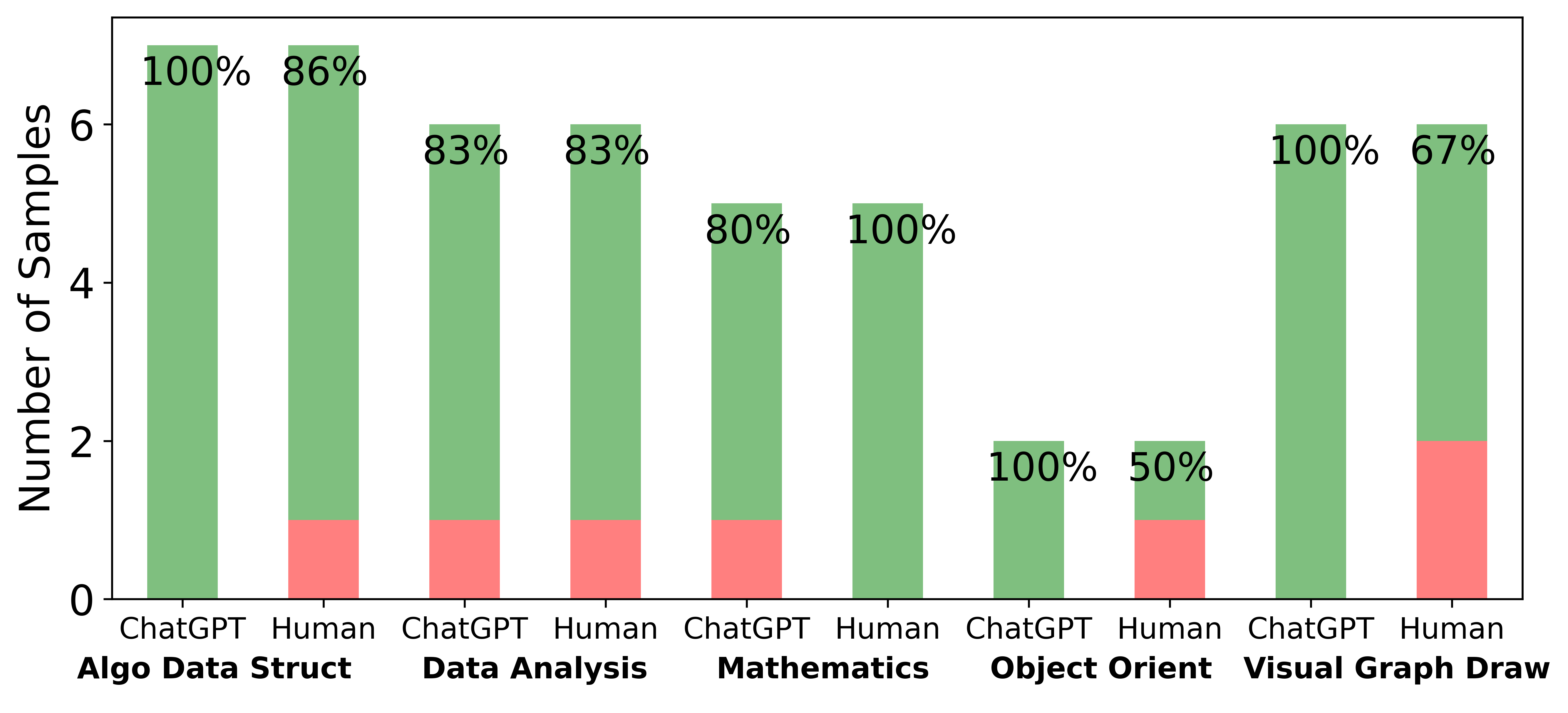}
    \caption{Random Forest}
    \label{fig:category_rf}
  \end{subfigure}
  \hfill
  \begin{subfigure}{0.32\linewidth}
    \includegraphics[width=\linewidth]{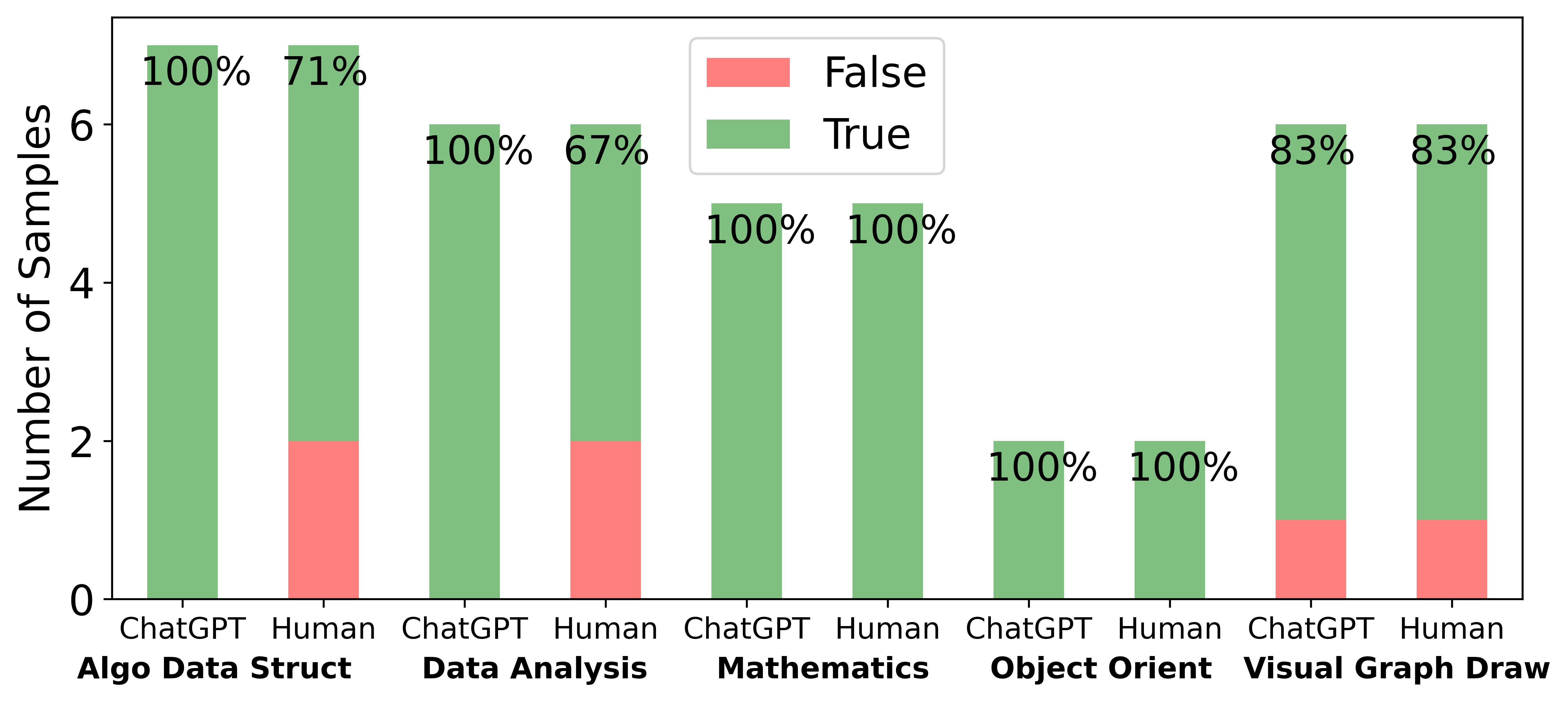}
    \caption{Decision Tree}
    \label{fig:category_dt}
  \end{subfigure}
  \hfill
  \begin{subfigure}{0.32\linewidth}
    \includegraphics[width=\linewidth]{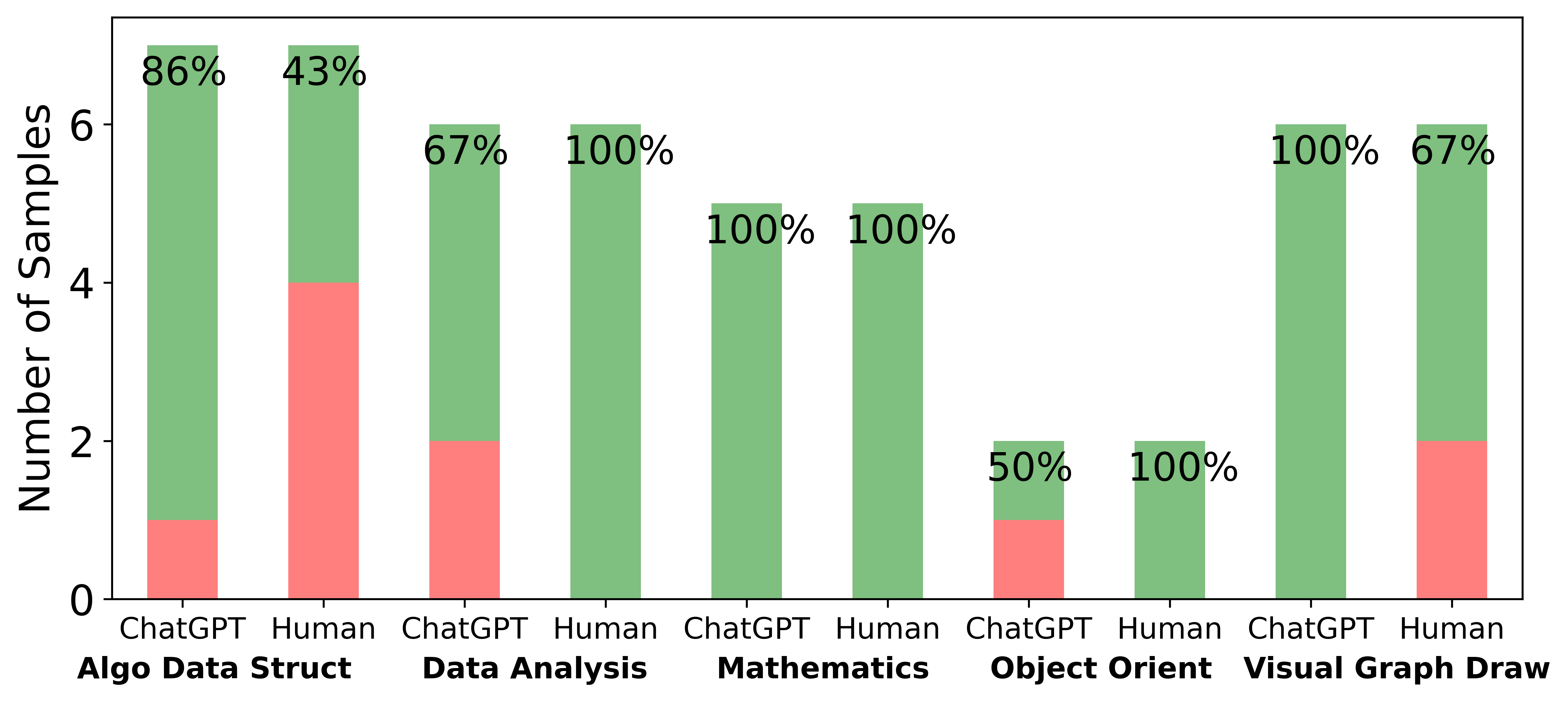}
    \caption{RUSBoost}
    \label{fig:category_rusboost}
  \end{subfigure}
  \vspace{-0.3cm}
  \caption{ML model performance for different categories (x-axis) in the test set}
  \label{fig:category_plots}
\end{figure}
  \ding{112}  \textbf{Per-category Performance Test.} Investigating the impact of varying train set ratios on model performance, we observed distinct behaviors among the three models, as illustrated in Figure~\ref{fig:category_plots}. The RF model exhibited remarkable stability, achieving the highest performance with an 87\% F1 Score at approximately an 80\% train split size. In contrast, the DT model demonstrated the highest accuracy at 88\% with an 80\% train split ratio. However, its performance exhibited inconsistency, fluctuating abruptly with an increasing split ratio. This variation may be attributed to the random sampling of the dataset after assigning the split ratio, resulting in the models being trained on different examples at each split. This phenomenon is akin to employing different cross-validation folds with varying split ratios. Notably, a significant performance decline was observed in the 40-50\% train split size range, where the F1 score plummeted to nearly 40\%. This decline suggests that the model performed poorly on the selected test set examples at that split ratio, indicating potential challenges in generalizability. Similarly, the performance of the RUSBoost model exhibited substantial variation as we increased the train set size. In contrast to RF, both DT and RUSBoost indicated a potential risk of overfitting the provided dataset, highlighting concerns about their generalizability. Interestingly, all models experienced a sharp drop in performance after reaching an 80\% train split size, likely due to the test set containing very few examples for the models to classify correctly.
   
\begin{figure}[ht]
  \centering
  \begin{subfigure}{0.30\linewidth}
    \includegraphics[width=\linewidth]{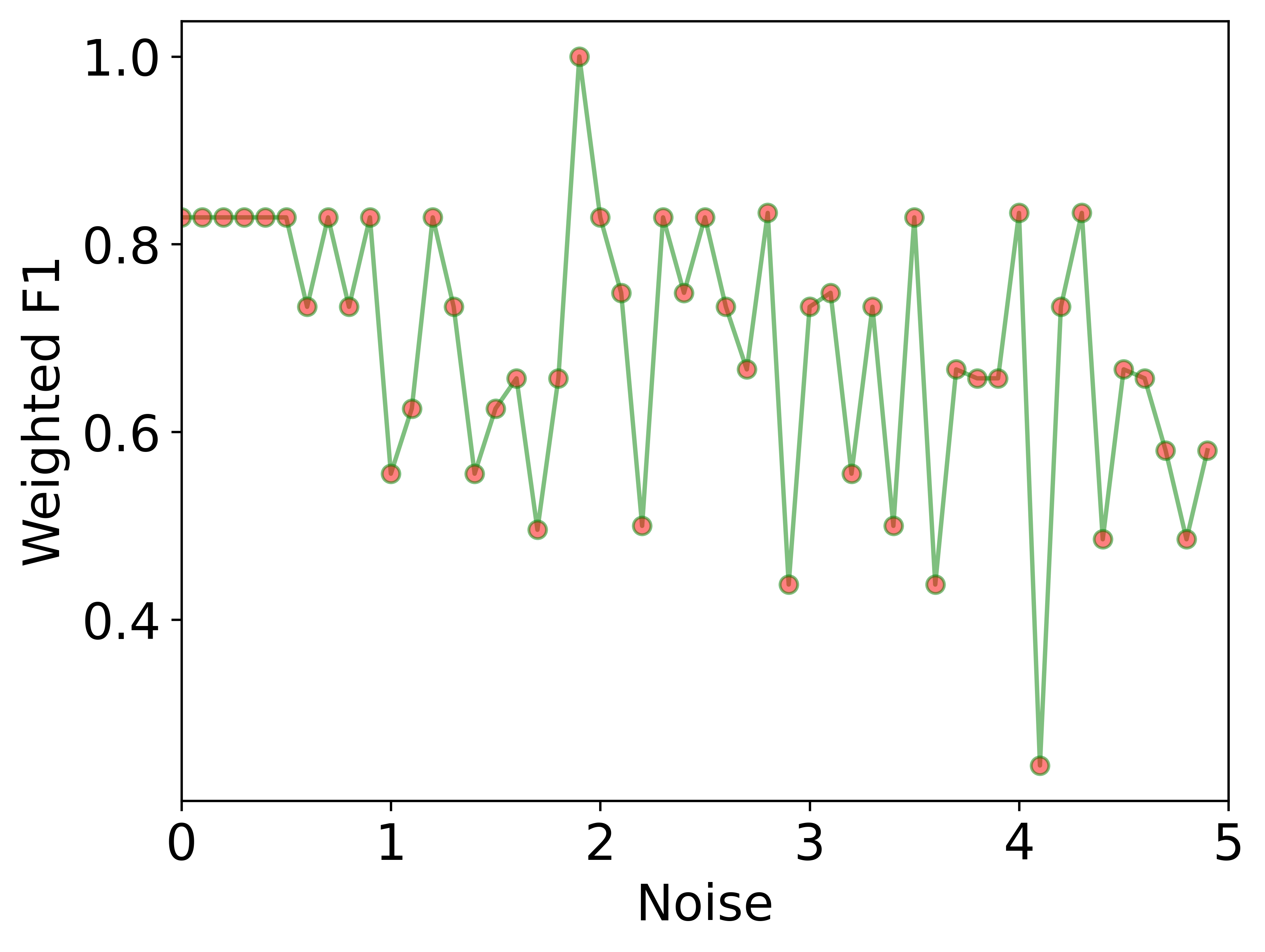}
    \caption{Random Forest}
    \label{fig:noise_rf}
  \end{subfigure}
  \hfill
  \begin{subfigure}{0.30\linewidth}
    \includegraphics[width=\linewidth]{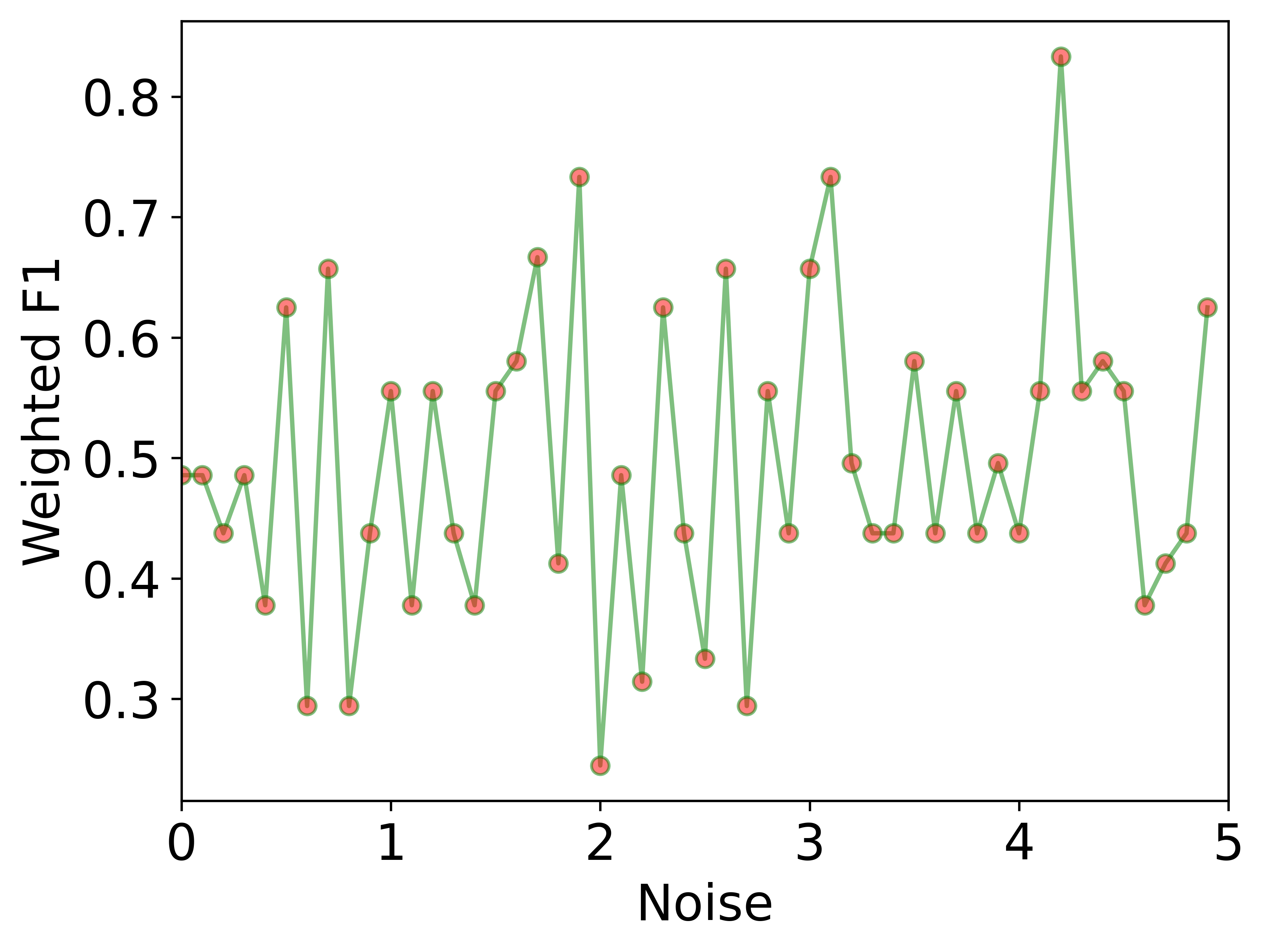}
    \caption{Decision Tree}
    \label{fig:noise_dt}
  \end{subfigure}
  \hfill
  \begin{subfigure}{0.30\linewidth}
    \includegraphics[width=\linewidth]{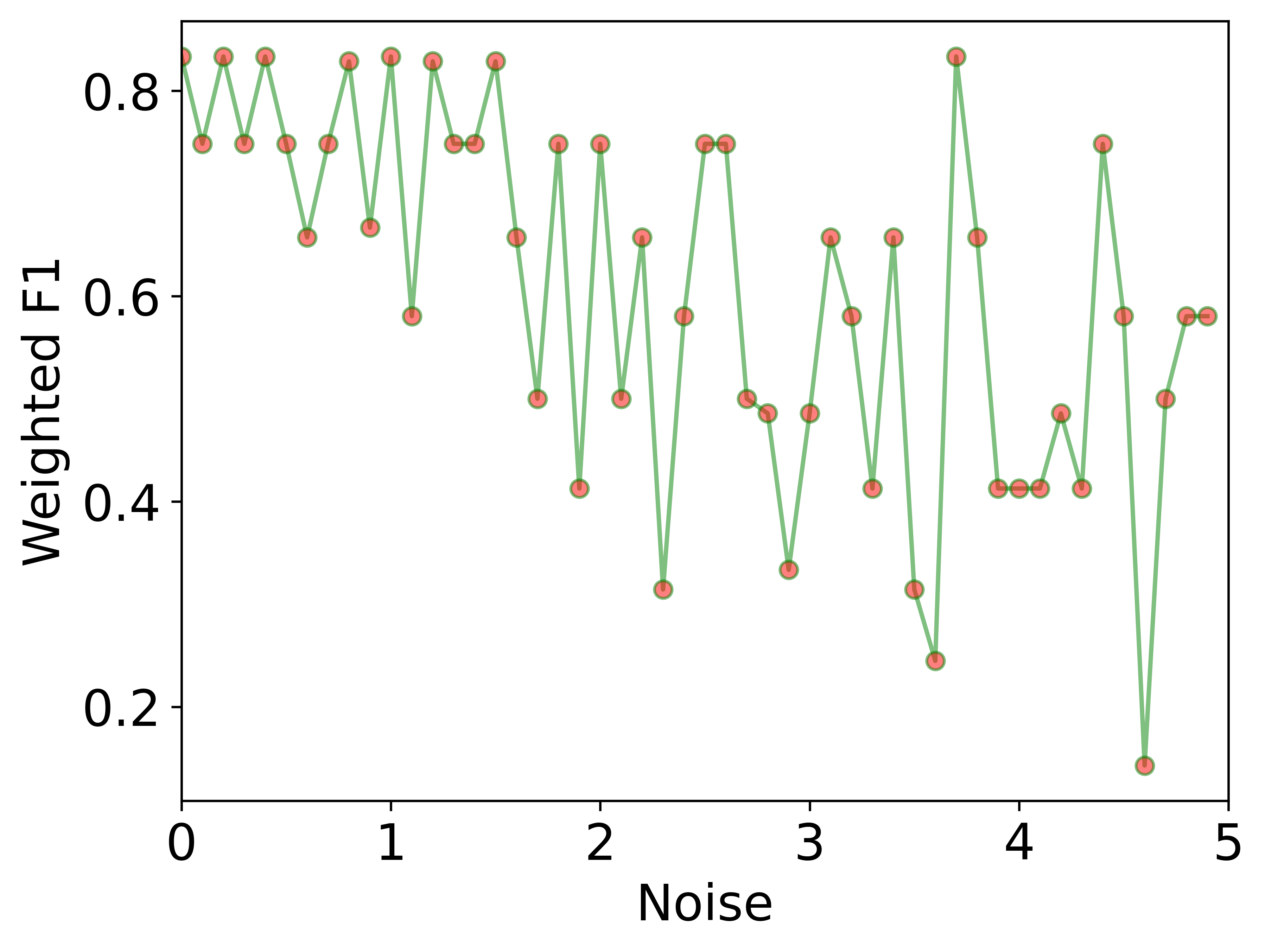}
    \caption{RUSBoost}
    \label{fig:noise_rusboost}
  \end{subfigure}
  \vspace{-0.4cm}
  \caption{ML model performance on the test set across the different amounts of Gaussian noise (x-axis)}
  \label{fig:noise_test}
\end{figure}
    \ding{112}  \textbf{Gaussian Noise Test.} Investigating model stability and generalizability, we introduced random noise to the test set, which is particularly valuable when models are trained on smaller datasets. In our Gaussian noise test, intriguing findings emerged. Random Forest (RF) models exhibited remarkable stability, with performance dropping only modestly from 88\% to approximately 83\% as noise was introduced--See Figure \ref{fig:noise_test}. Notably, the model maintained stability even with incremental increases in the standard deviation of the random Gaussian noise generator, ranging from 0 to 0.7. The inherent randomness of the added Gaussian noise led to fluctuations, reaching nearly 98\% when the standard deviation was increased to 2. While both RF and RUSBoost models displayed resistance to added noise, RF exhibited greater resilience in this Gaussian Noise Test, showcasing its comparative robustness.

Based on the results of the three tests, we conclude that the Random Forest (RF) algorithm shows robustness in terms of random sampling of changing train set size, addition of random Gaussian noise with varying standard deviation, and cross-category performance across the test set.

\subsection{Feature Analysis}
\vspace{-0.5cm}
\begin{table}[ht]
    \centering
    \caption{Feature Importance of the 14 coding metrics using the dual feature importance approach}
    \vspace{-0.2cm}
    \resizebox{0.65\linewidth}{!}{%
    \begin{tabular}{lccc} 
        \toprule
        \textbf{Feature} & \textbf{Forest Importance} & \textbf{Perm Importance} & \textbf{Mean Importance} \\
        \midrule
        Number of Comments & 0.172082 & 0.298360 & 0.235221 \\
        Number of Lines & 0.123837 & 0.179766 & 0.151802 \\
        Maintainability Index & 0.099583 & 0.178798 & 0.139190 \\
        Number of Functions & 0.088670 & 0.167553 & 0.128111 \\
        Logical SLoC & 0.085149 & 0.122872 & 0.104010 \\
        SLoC & 0.084966 & 0.122422 & 0.103694 \\
        Halstead Time & 0.049588 & 0.073684 & 0.061636 \\
        Halstead Volume & 0.048599 & 0.074152 & 0.061376	 \\
        Halstead Bugs & 0.048450 & 0.074172 & 0.061311	 \\
        Halstead Effort & 0.049134 & 0.072983 & 0.061058	 \\
        Cyclomatic Complexity & 0.048372 & 0.070053 & 0.059212 \\
        Halstead Difficulty & 0.045021 & 0.070368 & 0.057695 \\
        Diff SLoC LLoC & 0.045856 & 0.065013 & 0.055435 \\
        Number of Classes & 0.010693 & 0.014973 & 0.012833 \\
        \bottomrule
    \end{tabular}
    }
    \label{tab:feature_importance}
\end{table}

As depicted in Table~\ref{tab:feature_importance}, the \textit{Number of Comments} and \textit{Lines of Code} stand out as pivotal features, echoing the insights discussed in Section 5.1. While these metrics hold importance, particularly for RF models, our exploration delves deeper into other metrics of higher significance. Among these, the \textit{Maintainability Index} proves most paramount in predicting source code authorship. This index encompasses metrics such as \textit{Cyclomatic Complexity, Source Lines of Code}, and \textit{Halstead Volume}. Interestingly, the aggregate influence of the maintainability index surpassed the individual impacts of both \textit{Source Lines of Code} and \textit{Cyclomatic Complexity}. In terms of maintainability, ChatGPT-generated code consistently exhibited superior performance compared to human-written code. This superiority is likely attributable to its lower \textit{Halstead Volume}, reduced count of source lines, and minimized \textit{Cyclomatic Complexity}. On the lower end of the significance spectrum lie metrics such as \textit{Cyclomatic Complexity}, \textit{Number of Classes}, \textit{Difference between Logical and Source Lines of Code}, and \textit{Halstead Difficulty}. Their diminished importance might be a result of their strong correlation with dominant features like \textit{Source Lines of Code (SLoC)} and \textit{Logical SLoC}, making them somewhat redundant in the eyes of the algorithms, especially when other correlated metrics are present. A noteworthy point pertains to the \textit{Number of Classes} metric, ranking as the least important for determining authorship. The reason was both ChatGPT and human participants adhered to writing an identical number of classes, constrained by the specific requirements set in the prompts. The singular choice of Python as the language and the limitation to a single file size probably swayed this outcome. It suggests potential disparities if the prompts weren't bound to one file or if a language more inclined towards class utilization was chosen.

\section{Conclusion and Future Work} 

This paper presented a comprehensive evaluation of ChatGPT's code generation capabilities compared to human programmers. We curated a novel dataset of 131 prompts and analyzed 262 code solutions using quantitative metrics and qualitative techniques. Our findings reveal ChatGPT's strengths in efficiently generating concise, modular code with advanced constructs. However, limitations arose in visual-graphical challenges. The comparative analysis highlighted ChatGPT's superior error handling and maintainability while uncovering distinct coding style differences. Additionally, we developed highly accurate machine learning models to classify ChatGPT versus human code.

Our contributions establish a robust foundation to advance AI programming assistants. The curated dataset provides a benchmark for rigorous LLM evaluations. The quantitative and qualitative analyses deliver nuanced insights into current capabilities and limitations. Our ML models pioneer techniques to detect AI-generated code automatically. Several promising directions remain for future work. As LLMs continue evolving rapidly, prompt engineering techniques should be refined to enhance performance. Testing ChatGPT's abilities across diverse languages and integrating code context could reveal further insights. Evaluating code generation alongside complementary tasks like summarization and documentation may be worthwhile. Investigating neural architecture optimizations and continual learning to bolster visual reasoning could address key weaknesses. Overall, this work helps chart a course toward trustworthy and human-aligned AI programming assistants.

\bibliographystyle{ACM-Reference-Format}
\bibliography{References}

\end{document}